\documentclass[longbib,PRB,reprint,superscriptaddress]{revtex4-1}
\usepackage{marvosym}
\usepackage{graphicx}
\usepackage{lipsum}
\usepackage{amssymb}
\usepackage{bbold}
\usepackage{amsmath}
\usepackage{bm}
\usepackage{url}
\usepackage{prettyref}
	\newcommand{\pr}[1]{\prettyref{#1}}
	\newrefformat{app}{appendix~\ref{#1}}
	\newrefformat{eqn}{Eqn.~(\ref{#1})}
	\newrefformat{eqns}{Eqns.~(\ref{#1})}
	\newrefformat{sec}{section~\ref{#1}}
	\newrefformat{subsec}{section~\ref{#1}}
	\newrefformat{subsubsec}{section~\ref{#1}}
	\newrefformat{fig}{Fig.~\ref{#1}}

\newcommand{\V}[1]{\mathbf {#1}}
\usepackage[usenames,dvipsnames]{color}

\newcommand{\notes}[1]{{\color{PineGreen}(\url{#1})}}
\usepackage[breaklinks,bookmarksnumbered]{hyperref}
\begin{document}
\title{Faraday effect in rippled graphene: Magneto-optics and random gauge fields}
\author{J{\"u}rgen Schiefele}
\affiliation{%
Instituto de Ciencia de Materiales de Madrid,
CSIC,
E-28\,049 Madrid, Spain
}
\author{Luis Martin-Moreno}
\affiliation{%
Instituto de Ciencia de Materiales de Arag\'on 
and
Departamento de F\'isica de la Materia Condensada, 
CSIC-Universidad de Zaragoza, 
E-50\,009 Zaragoza, Spain
}
\author{Francisco Guinea}
\affiliation{%
IMDEA Nanociencia, Calle de Faraday~9,
E-28\,049 Madrid, Spain}
\affiliation{%
Department of Physics and Astronomy, University of Manchester, Oxford Road, Manchester M13~9PL, United Kingdom
}
\date{\today}
\begin{abstract}
A beam of linearly polarized light transmitted through magnetically biased graphene can have its axis of
polarization rotated by several degrees after passing the graphene sheet.
This large Faraday effect is due to the action of the magnetic field on graphene's charge carriers. 
As deformations of the graphene
membrane result in pseudomagnetic fields acting on the charge carriers,
the effect of random mesoscopic corrugations (ripples)  
can be described as the exposure of graphene to a random
pseudomagnetic field.
We aim to clarify the interplay of these typically sample inherent fields with the  external magnetic bias
field and the resulting effect on the Faraday rotation.
In principle, random gauge disorder can be identified from a combination of Faraday angle and optical spectroscopy measurements.
\end{abstract}
\maketitle
\section{Introduction}
Elastic scattering of light by a conducting medium, and thus the reflection and
absorption of a beam of light passing through a slab of material, are largely
determined by the interaction of the beam with the material's free electrons.
If an external magnetic field is applied to the medium, the Lorentz force acts
on the conduction electrons, and if, for definiteness, the electric component
of the incident beam is linearly polarized along the $x$-direction (see \pr{fig:sketch}), the transmitted as
well as the reflected beam both acquire a component along the $y$-direction.  This
rotation of the polarization direction in the reflected (transmitted) beam is
called the Kerr (Faraday) effect.
The direction of the rotation depends only on the direction of the magnetic field, 
not on the propagation direction of light, such that Faraday and Kerr effect enable non-reciprocal
optical devices\cite{Chin_2013,Peng_2014b,Bi_2011} like optical isolators, which allow light to pass in one direction while blocking
propagation in the reverse direction.

\begin{figure}
\centering
\includegraphics[width=0.85\columnwidth]{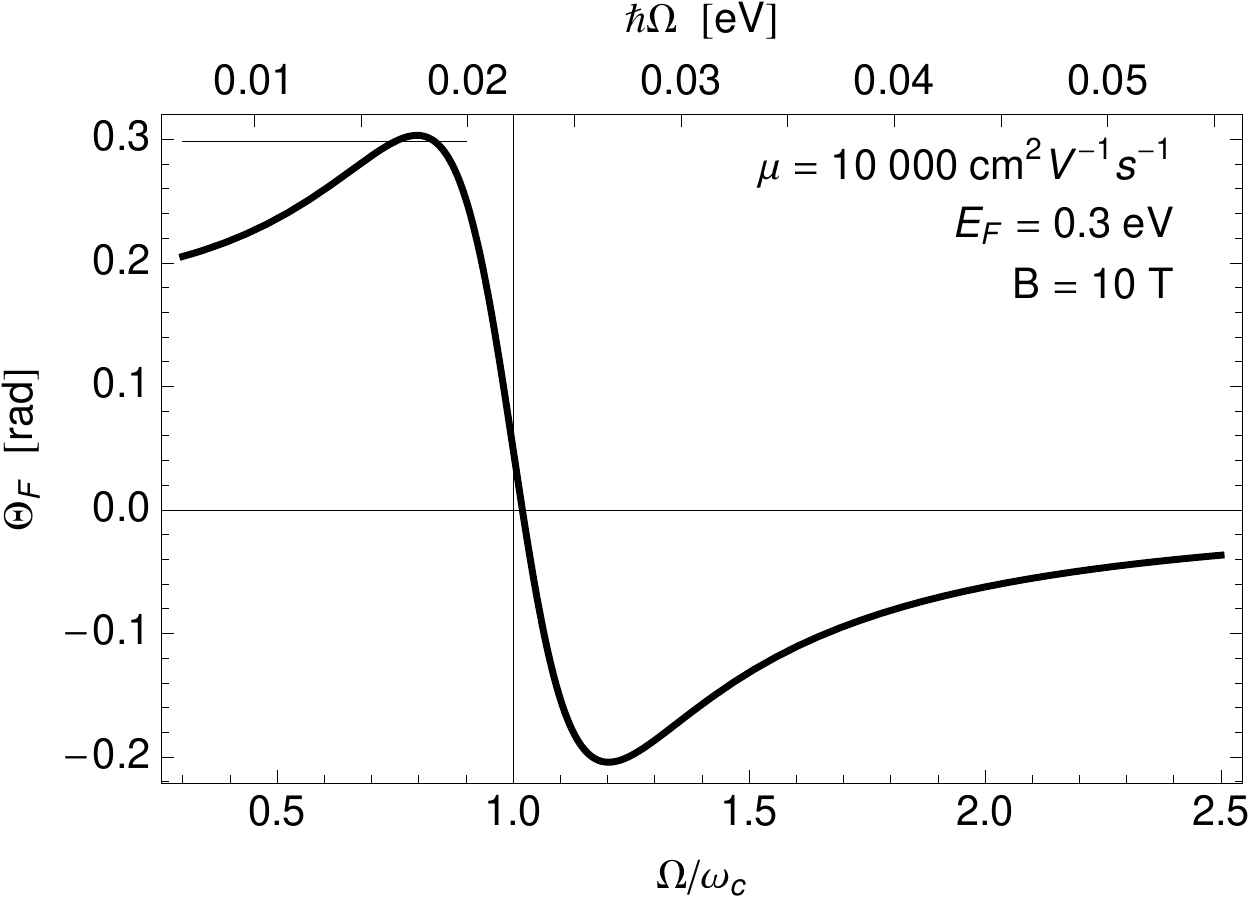}
\caption{%
Faraday rotation angle $\Theta_F$ versus frequency,
calculated  from the classical Drude formula
\pr{eqn:sxy_Drude_int} for $\sigma_{xy}$.
The horizontal line indicates the
maximal value of $\Theta_F$ [see \pr{eqn:max}] at frequencies below the cyclotron
frequency $\omega_c$. The scattering time has been set to
$\tau$=$\tau_{\rm tr}$=1.5$\times$10$^{-13}$\,s.
\label{fig:classicPlot}
}
\end{figure}

For the particular case of a beam of light of frequency $\Omega$ passing
through an atomically thin sheet of monolayer graphene in a 
constant magnetic field of magnitude $B$, one finds that the polarization is rotated by the
Faraday angle\cite{Fialkovsky_2012}
\begin{align}
\Theta_F
	\simeq
	\frac 1 2 \operatorname{Re} \sigma_{xy}(\Omega,B)
.
\label{eqn:TF_def}
\end{align}
Under typical experimental conditions, monolayer graphene can rotate the polarization by several degrees\footnote{%
We use units $\epsilon_0$=$c$=1, such that $e^2/h$=$2\alpha$=$\frac 2 {137}$ radians.
In the typical experimental setting (see for example Ref.~\protect\cite{Crassee_2011}), graphene is deposited on a dielectric substrate.
Depending on the refractive index of the specific substrate material used, the prefactor $\frac{1}{2}$ in  Eqn.~(\protect\ref{eqn:TF_def})
is modified, such that the rotation angles reached with 
graphene-on-SiC devices are by a factor of $\simeq$1.7 smaller than those we show for free-standing 
graphene in Fig.~\protect\ref{fig:classicPlot}.
},
a large effect considering that Faraday isolators based on conventional magneto-optical materials
require propagation distances of several millimeters\cite{Chin_2013}.
\pr{eqn:TF_def} neglects terms of higher than linear order in the fine-structure constant,
and it is assumed that the magnetic field as well as the propagation direction of the
incident light are both perpendicular to the graphene surface.
$\sigma_{xy}(\Omega,B)$ denotes the off-diagonal element of graphene's conductivity-tensor,
such that a measurement of $\Theta_F$ represents an optical (that is, $\Omega\neq0$)  
analogue to a dc measurement of the Hall current.

\begin{figure}
\centering
\includegraphics[width=0.6\columnwidth]{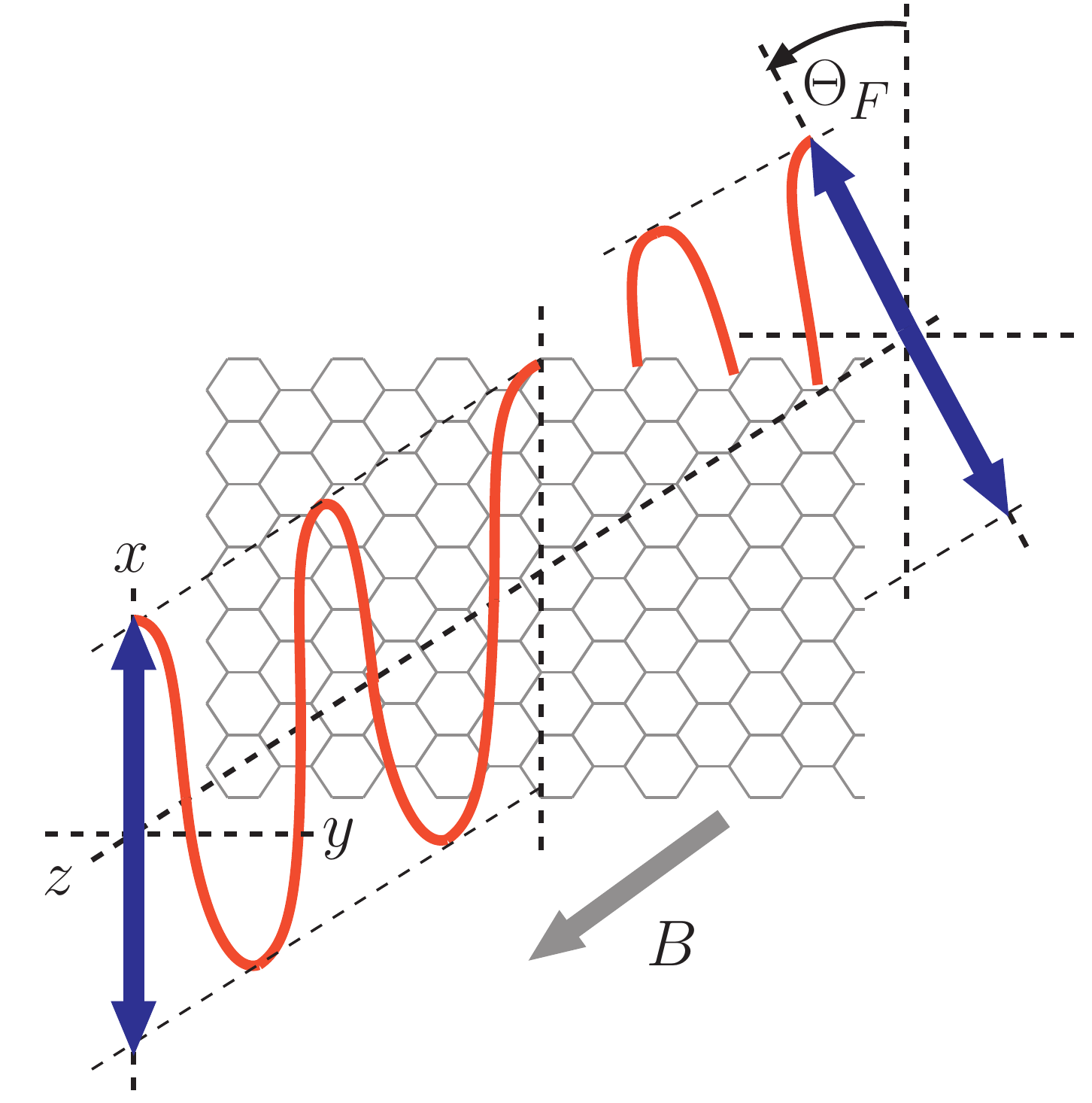}
\caption{%
The Faraday effect in graphene: The polarization plane of a linearly polarized
incoming beam is rotated by the Faraday angle $\Theta_F$ after passing through graphene
in a perpendicular magnetic field.
\label{fig:sketch}
}
\end{figure}

In the static limit, the dc Hall conductivity of graphene is known to show a step-like
behavior as a function of the filling factor $n_0$ \cite{Guinea_2009,Novoselov_2005,Zhang_2005},
\begin{align}
\label{eqn:sxyDC}
\sigma_{xy}(\Omega=0)
	&=\frac{e^2}{h} \;2 (1+2 n_0)
,
\end{align}
where $n_0$ marks the last filled level right below the Fermi energy $E_F$. 
For low carrier densities and  large magnetic fields, the Landau levels around $E_F$ are
well separated, and 
correspondingly, a Faraday angle displaying quantized steps as in \pr{eqn:sxyDC}
has been observed at $\Omega$=1\,THz for graphene samples with $E_F$=60\,meV and $B$ up to 7\,T \cite{Shimano_2013}.

However, the spacing of graphene's Landau levels $E_n$ 
is non-equidistant\cite{Guinea_2009}, with the energy difference between adjacent levels getting  smaller for higher $n$. 
Correspondingly, for modest magnetic fields, and with carrier densities in typical graphene-on-substrate samples being far from the neutrality point, 
the experimentally observed $\Theta_F(\Omega)$\cite{Crassee_2011}  follows the classical  off-diagonal
Drude conductivity\cite{Ashcroft_Mermin,Sounas_2011,Tymchenko_2013}, 
\begin{align}
\sigma_{xy}(\Omega)
	&=
	- \operatorname{sgn}(e B) \frac{e^2}{2 h} 
	\frac{ g_s g_v \omega_c E_F }{\hbar[(\Omega +  i/  \tau)^2 - \omega_c^2]}
,
\label{eqn:sxy_Drude_int}
\end{align}
which does not resolve the quantized Landau levels. 
\pr{eqn:sxy_Drude_int} is parametrized in terms of a cyclotron frequency\footnote{%
If we assume that the Lorentz force results in a circular
electronic motion with angular frequency $\omega_c$, then 
$
m^* \ddot{\V{r}}
	=
	-e \dot{\V{r}} \times \V{B}/c
$
with the effective electron mass $m^*=|E_F|/v_F^2$ (see Ref.~\protect\cite{Guinea_2009}) leads to \pr{eqn:wc_def}.
}
\begin{align}
\omega_c
	&=
	\frac{|e B| v_F^2}{|E_F|}
,
\label{eqn:wc_def}
\end{align}
$v_F$ denotes the Fermi velocity, $g_s$=$g_v$=2 are factors for the spin- and valley degeneracy,
and  $\tau$ is a phenomenological scattering time usually employed as a fitting parameter \cite{Crassee_2011}.

\pr{fig:classicPlot} shows $\Theta_F$ versus frequency calculated with \pr{eqn:sxy_Drude_int} for typical
values of $E_F$ and  $B$, and assuming $\tau$=1.5$\times$10$^{-13}$\,s, the value corresponding to the 
transport scattering time $\tau_{\rm tr}=\hbar \mu v_F/e$ for  a graphene sample with dc mobility $\mu$=10\,000\,cm$^2$V$^{-1}$s$^{-1}$\cite{Hwang_2008b}.
The rotation angle is seen to be maximal at frequencies slightly below $\omega_c$.
Assuming $\omega_c\tau\gg1$, which allows for closed cyclotron orbits between collisions,
\pr{eqn:sxy_Drude_int} yields 
\begin{align}
\Theta_F^{\rm max}
	&=
	\frac{e^2}{h}
	\frac{E_F}{4 \hbar\omega_c}
	\bigl\{\omega_c\tau + 1 +\mathcal{O}\bigl[1/(\omega_c\tau)\bigr]\bigr\}
\label{eqn:max}
\end{align}
for this maximal value of $\Theta_F$.
From \pr{eqn:max}, it is clear that the magnitude of the achievable Faraday rotation 
strongly depends on $\tau$, that is, on the quality of the specific graphene sample.

Due to the particular structure of graphene's two-dimensional crystal lattice, 
which results in a two-component spinor structure of the electron wavefunctions,
it is not only the magnitude of disorder that determines carrier dynamics. Rather, 
a microscopic description of disorder effects in graphene has 
to distinguish between scalar- and gauge potentials\cite{Guinea_2009}.
While the former are typically caused by charged impurities, the latter arise due to deformations
of graphene's crystal lattice,
and their effect on the charge carriers can be described in terms of pseudomagnetic fields\cite{Guinea_2008,Vozmediano_2010,Amorim_2015}.
Pseudomagnetic fields
affect carriers near the two inequivalent Fermi points $\V K$ and $\V{K}'$  with an opposite sign,
such that they either add up to or have to be subtracted from the (real) external bias field $B$ acting on the carriers\cite{Roy_2013b,Low_2010}.
As it has been found that the strain-induced pseudomagnetic field in graphene nano-bubbles can lead to a Landau quantization
of the charge carriers  equivalent to field strengths of a few hundred Tesla\cite{Levy_2010},
it seems possible to engineer the magneto-optical response of graphene via the controlled application of non-homogeneous
mechanical strain, if a constant pseudomagnetic field can be produced over a large spatial region\cite{Guinea_2010b,Zhu_2015}.

Apart from such engineered strain distributions, random strain fields are present in most available graphene samples, because graphene membranes, 
whether supported by a substrate material or in  suspended devices, are never completely flat.
Instead, they show a rippled structure\cite{Meyer_2007},
and for clean samples, 
the pseudomagnetic fields caused by these mesoscopic corrugations have been shown to 
set the dominant limit on electronic transport\cite{Couto_2014}.
Despite this importance of ripples for graphene's transport properties, experiments on the Faraday effect in graphene do not reveal any 
valley-dependent interplay between (potentially large) gauge fields and the magnetic bias field. 
Instead, the data is found to be fitted well by \pr{eqn:sxy_Drude_int} with an
appropriately chosen scattering time
\cite{Crassee_2011,Shimano_2013,Fialkovsky_2012}.

It is the aim of the present work to investigate the effect of sample-inherent, random strain configurations on the magneto-optical properties of graphene,
and to understand why ripple-induced effective magnetic fields of the form $B\pm\Delta B$ in each valley do not seem to 
play a role in graphene's Faraday effect.
To this end, we compare the effect of random ripples on the Faraday rotation
with that of random scalar scatterers.
\section{Disorder scattering in graphene -- gauge and scalar potentials}
\label{sec:SE}
In order to relate the scattering parameter $\tau$ appearing in \pr{eqn:sxy_Drude_int} to an underlying  microscopic scattering mechanism,
we assume random Gaussian disorder\cite{Akkermans_Montambaux}. 
Denoting the Fourier transform of the scattering potential with $V_{\V q}$, 
disorder scattering  can be introduced in perturbation theory by 
the correlator $\langle V_{\V q} \otimes V_{-\V q} \rangle_{\rm dis}$,
where the average is taken over all possible disorder configurations.
As the electron-wavefunction $\Psi$ is a spinor of rank four in sublattice (superscripts $A$ and $B$)
and valley-space (subscripts $\V{K}$ and $\V{K}'$) \footnote{%
See Ref.~\protect\cite{Guinea_2009}. We here neglect the electronic spin degree of freedom.}%
, 
\begin{align}
\Psi
	&=
	\{
	\psi^A_{\V K}, \psi^B_{\V K},  \psi^A_{{\V K}'},  \psi^B_{{\V K}'}  
	\}
,
\nonumber
\end{align}
the correlator forms a tensor 
\begin{align}
\bigl[
\langle V_{\V q} \otimes V_{-\V q} \rangle_{\rm dis}
\bigr]_{h i; j k}
,\qquad h,i,j,k \in \{1,2,3,4\}
,
\label{eqn:corr}
\end{align}
see \pr{fig:K0}b.

We want to distinguish between a scalar and a gauge potential $V^{s,g}$, 
and consider
\begin{subequations}
\label{eqns:corr_match}
\begin{align}
\langle V^s_{\V q} \otimes V^s_{-\V q} \rangle_{\rm dis}
	&=
	g \; \mathbb{1} \otimes \mathbb{1},
\label{eqn:corr_match_s}
\\
\langle V^g_{\V q} \otimes V^g_{-\V q} \rangle_{\rm dis}
	&=
	g \; \gamma_1 \otimes \gamma_1, 
\label{eqn:corr_match_g}
\end{align}
\end{subequations}
where $\mathbb{1}$ and $\gamma_1$ are 4$\times$4 matrices.
The scalar potential $V_{\V q}^s$ affects carriers on the $A$ and $B$ sublattices of
graphene and in both $\V K$ and $\V{K}'$ valleys in the same manner.
In contrast, the gauge potential is proportional to $\gamma_1$, defined as
\begin{align}
\gamma_i
	&
	=
	\left( \begin{array}{cc}
	\sigma_i&0\\
	0&-\sigma_i\end{array}\right)
,
\nonumber
\end{align}
where we use the standard 
2$\times$2 Pauli matrices  $\sigma_{i}$ acting on the sublattice index, that is,  $A$ or $B$ site \cite{Guinea_2009}. 
Thus, $V_{\V q}^g$ affects carriers in each valley with opposite sign.

Apart from their different tensor structure, we have set both correlators proportional
to a  constant $g$ of dimensions (energy)$^2$$\times$area,
as would be the case for sharply localized, $\delta$-like scattering centers.
In a more detailed description, gauge as well as scalar disorder potentials are likely to show long-range correlations,
resulting in an overall power-law dependence of the correlator in \pr{eqn:corr_match_g} \cite{Couto_2014,Guinea_2008}. 
As we will see below,  the approximation with a momentum-independent constant $g$ greatly 
facilitates the evaluation of otherwise infrared-divergent loop integrals,
while still taking into full account the scalar or gauge nature of the disorder.
Further, the block-diagonal potentials in  \pr{eqns:corr_match} neglect possible inter-valley scattering processes, which have been shown to yield only weak contributions
to the resistivity of graphene\cite{Couto_2014}.

In order to assign a value to the disorder strength $g$, we aim to link it to the dc carrier mobility characteristic for each sample.
We assume the dc conductivity of graphene to be of the Drude-form
$\sigma_{xx}^{\rm dc}
	=
	\frac{e^2}{h} { 2 v_F k_F\tau_{\rm tr}}$,
where $\tau_{\rm tr}$ is obtained from Fermi's rule together 
with a Boltzmann weighting factor $1-{\rm cos} \phi_\mathbf{q}$
which supresses contributions of small-angle scattering events (${\rm cos} \phi_\mathbf{q}\simeq1$) to the dc conductivity\cite{Hwang_2008b,Couto_2014}.
Then, the definition of $\mu$ 
as the electrical conductivity per carrier, $\mu=\sigma^{\rm dc}_{xx}/(n e)$, where $n$
is the carrier density,
leads to
\begin{align}
\mu
	&=
	\frac{e}{h} \frac{2 \pi v_F \tau_{\rm tr}}{k_F}
.
\label{eqn:mu_relation}
\end{align}
Transport measurements revealed that $\mu$ is independent of the carrier density\cite{Couto_2014}, hence the
scattering parameter $\tau_{\rm tr}$ has to be proportional to $k_F$ and, as a function of $\mu$, has to fulfill
the above relation.

The retarded electronic self energy due to scattering at the potential $V^{s}$ or  $V^{g}$ in Born approximation 
(see the diagram in \pr{fig:K0}a)
is given by
\begin{align}
\Sigma^{s,g}(\Omega)
	&=
	g \;(\hbar \Omega +E_F)
\nonumber
\\
	&\times
	\int_0^\infty \frac{q dq}{2 \pi}
	\frac{1}{(\hbar \Omega +E_F +i 0)^2-(\hbar v_F q)^2}
,
\label{eqn:SE_born}
\end{align}
Importantly, it does not distinguish between a scalar or vector character of the underlying disorder.
The imaginary part of $\Sigma(\Omega)$ describes the broadening of electronic momentum-eigenstates
due to the disorder potential, and defines a quantum scattering time $\tau_q$,
\begin{align}
\frac 1 {\tau_q}
	&=
	-\frac{2}\hbar \operatorname{Im} \Sigma \big{|}_{\Omega=0}
	=
	g \, \frac{k_F}{2\hbar^2 v_F}
.
\label{eqn:tau_of_mu}
\end{align}
For white-noise disorder, $\tau_q$ is known to be smaller than $\tau_{\rm tr}$ by a factor of 2 \cite{Hwang_2008b},
and form \pr{eqn:tau_of_mu} together with the requirement \pr{eqn:mu_relation}, we arrive at 
\begin{align}
g
	&=
	4\pi \frac{e}{h \mu} \frac{(\hbar v_F)^2}{k_F^2}
\label{eqn:gdef}
\end{align}
for the disorder strength in a graphene sample with mobility $\mu$.

\section{Disorder corrections to the Hall conductivity}
\label{sec:simple_bubble}
From Kubo's formula, the Hall conductivity of graphene can be expressed as\cite{Gusynin_2007c,Gusynin_2006}
\begin{align}
\sigma_{xy}(\Omega)
	&=
	\frac{e^2}{h} \operatorname{Im}	\frac{4 \pi v_F^2}{\Omega}
	\int d\omega d\omega'
	\frac{n_F(\omega')-n_F(\omega)}{\omega-\omega'-\Omega-i 0}
\nonumber
\\
	&\times
	\sum_{n=0}^\infty K_n^{(0)}(\omega,\omega')
,
\label{eqn:sxy_bare}
\end{align}
with $n_F$ the Fermi distribution, and  the integral kernel $K_n^{(0)}$
resulting from a sum over bubble diagrams, see \pr{fig:K0}c:
\begin{align}
&
\sum_{n=0}^\infty K_n^{(0)}(\omega,\omega')
\label{eqn:Kubo_Kernel_clean}
\\
	&=
	\sum_{\beta,\gamma=\pm1}
	\frac{\beta \gamma}{(2 \pi i)^2}
	\int\frac{d^2 k}{(2\pi)^2}
	\operatorname{Tr}\bigl[
	\gamma_1 G^\beta(\omega, \V k) \gamma_2 G^\gamma(\omega',\V k)
	\bigr]
\nonumber
\end{align}
The sum over the indices $\beta$ and $\gamma$ realizes different combinations
of retarded and advanced Greens functions $G^{\pm1}$ for charge carriers,  given in \pr{eqn:Gdef}.

Disorder corrections can enter \pr{eqn:Kubo_Kernel_clean} either through self-energy insertions (as shown in \pr{fig:K0}a), 
which shift the resonances of the Greens functions in the bubble diagrams of \pr{fig:K0}c, or through vertex corrections,
shown in \pr{fig:K0}d, see Ref.~\cite{Briskot_2013}.
\subsection{Self-energy insertions}
\label{subsec:bubble}
Within the lowest-order Born approximation, we assume the real part of disorder-induced self-energies
to be negligible, such that they simply introduce 
the scattering time $\tau_q$ of \pr{eqn:tau_of_mu} 
into the Greens functions in \pr{eqn:Kubo_Kernel_clean}.
For such a constant damping parameter, the bubble diagrams of \pr{eqn:sxy_bare} have been evaluated in
Refs.~\cite{Gusynin_2006,Gusynin_2007c},
and we only recall the main steps of the calculation.

After performing the momentum integration in \pr{eqn:Kubo_Kernel_clean} with the help of relation (\ref{eqn:otho_L}),
$K_n^{(0)}$ can be expressed in terms of the polarization functions
\begin{align}
\Pi^{\beta \gamma}_{m n}(\omega,\omega')
	&=
	\frac{2 (\hbar v_F)^2}{l_B^2}
	\frac
	{\hbar \omega + i \beta \hbar/(2\tau_q) }
	{\hbar^2[\omega + i \beta /(2\tau_q)]^2 - E_n^2 }
\nonumber
\\
	&\times
	\frac
	{\hbar\omega' + i \gamma \hbar/(2\tau_q)}
	{\hbar^2[\omega' + i \gamma /(2\tau_q)]^2 - E_m^2  }
,
\label{eqn:Pidef}
\end{align}
where the Landau levels are given by
\begin{align}
E_n
	&=
	\operatorname{sign}(n) \, \hbar v_F \sqrt{2 |n|}/l_B
,
\label{eqn:LL_def}
\end{align}
with $l_B = \sqrt{\hbar/|e B|}$ the magnetic length. 
We arrive at\cite{Gusynin_2006} 
\begin{align}
&
K_n^{(0)}(\omega,\omega')
	=
	i\operatorname{sgn}(e B) \frac{ g_s g_v}{ 4 (2 \pi)^3 v_F^2}
\nonumber
\\
	&\times
	\sum_{\beta,\gamma=\pm1}
	\beta\gamma
	\bigl[
	\Pi^{\beta \gamma}_{n+1, n}(\omega,\omega')
	-
	\Pi^{\beta \gamma}_{n, n+1}(\omega,\omega')
	\bigr]
,
\label{eqn:Kubo_Kernel_clean2}
\end{align}
$g_s$=$g_v$=2 denoting spin and valley degeneracy, respectively.

\begin{figure}
\centering
\includegraphics[width=0.80\columnwidth]{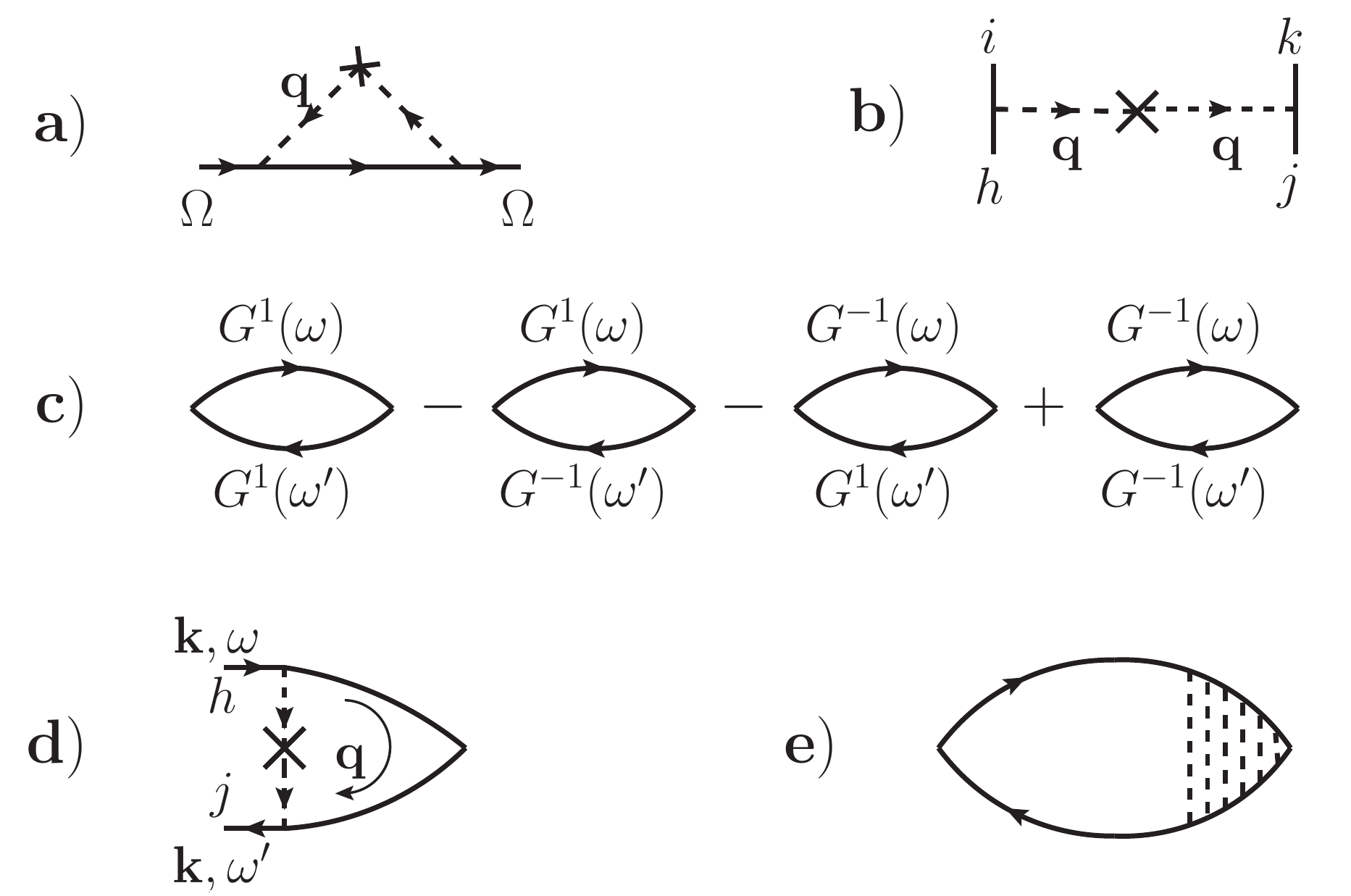}
\caption{%
\label{fig:K0}
Diagrammatic expressions appearing in the calculation of $\sigma_{xy}$.
{\bf a)}~Self energy $\Sigma(\Omega)$, see \pr{eqn:SE_born}.
{\bf b)}~Disorder correlator $\langle V_{\V q} \otimes V_{-\V q} \rangle_{\rm dis}$, see \pr{eqn:corr}.
{\bf c)}~Integral kernel of \pr{eqn:Kubo_Kernel_clean}.
{\bf d)}~Vertex correction $\Gamma(\V k, \omega,\omega')$, see \pr{eqn:vertex_correction_general}.
{\bf e)}~Resummation of vertex corrections leading to Eqn.~(\ref{eqn:sxy_resum_scalar_3}).
}
\end{figure}
\subsection{Vertex corrections}
\label{subsec:vertex}
Apart from self-energy corrections contributing to the scattering time,
disorder introduces, in leading order of perturbation theory, a vertex correction .
\begin{align}
&
\bigl[ \Gamma^{\beta\gamma}(\V k, \omega,\omega')\bigr]_{h j}
	=
	\int \frac{d^2q}{(2\pi)^2} \, 
	\bigl[ \langle V_{\V q - \V k} \otimes V_{-\V q + \V k} \rangle_{\rm dis}\bigr]_{h i;j k}
\nonumber
\\
	&\times
	\bigl[G^{\beta} (\omega,\V q)
	\gamma_2
	G^{\gamma} (\omega',\V q)\bigr]_{i k}
\label{eqn:vertex_correction_general}
\end{align}
(see \pr{fig:K0}d), which  replaces the bare velocity operator $\gamma_2$ appearing  
on the right-hand-side of \pr{eqn:Kubo_Kernel_clean}.
For $V_{\V q}$ a constant independent of $\V q$, the integration
over the loop momentum in \pr{eqn:vertex_correction_general} can be performed using \pr{eqn:otho_L}, and for the scalar and gauge
disorder of \pr{eqns:corr_match}, we obtain
\begin{align}
\Gamma^{\beta\gamma}_{s,g}(\V k, \omega,\omega')
	&=
	\frac{g}{4 \pi (\hbar v_F)^2}
	\gamma_2 \cdot \operatorname{diag}(a^{\beta\gamma}_{s,g},b^{\beta\gamma}_{s,g},a^{\beta\gamma}_{s,g},b^{\beta\gamma}_{s,g})
,
\label{eqn:vertex_correction2}
\end{align}
where the entries $a$ and $b$ of the diagonal matrix are dimensionless functions of the frequencies
$\omega$ and $\omega'$:
\begin{subequations}
\label{eqns:ab_strain_loop}
\begin{align}
a^{\beta \gamma}_s(\omega,\omega')	
	&=
b^{\beta \gamma}_g(\omega,\omega')	
	=
	\sum_{\overline{n}=0}^\infty
	\Pi_{\overline{n}, \overline{n}+1}^{\beta \gamma}(\omega,\omega')	
,
\label{eqn:b_strain_loop}
\\
b^{\beta \gamma}_s(\omega,\omega')	
	&=
a^{\beta \gamma}_g(\omega,\omega')	
	=
	\sum_{\overline{n}=0}^\infty
	\Pi_{\overline{n}+1, \overline{n}}^{\beta \gamma}(\omega,\omega')	
,
\label{eqn:a_strain_loop}
\end{align}
\end{subequations}
such that $\Gamma_g=\gamma_1\Gamma_s\gamma_1$.

Resumming the vertex correction \pr{eqn:vertex_correction2} in a geometric series  as indicated in
\pr{fig:K0}e yields, 
for either scalar (superscript $s$) or gauge (superscript $g$) disorder,
\label{eqn:sxy_resum_scalar}
\begin{align}
\sigma_{xy}^{s,g}(\Omega)
	&=
	\frac{e^2}{h} \operatorname{Im}	\frac{2 (\hbar v_F)^2}{\pi \Omega}
	\int d\omega d\omega'
	\frac{n_F(\omega')-n_F(\omega)}{\omega-\omega'-\Omega-i 0}
\nonumber
\\
	&\times
	{i \operatorname{sign}(e B)}
	\sum_{\gamma,\beta=\pm1}
	K_{s,g}^{\beta\gamma}(\omega,\omega')
.
\label{eqn:sxy_resum_scalar_3}
\end{align}
Here
\begin{subequations}
\label{eqns:Kres}
\begin{align}
K_s^{\beta\gamma}(\omega,\omega')
	&=
	\frac{b_s^{\beta\gamma}(\omega,\omega')}{{4\pi(\hbar v_F)^2 }-{g}\,b_s^{\beta\gamma}(\omega,\omega')}
\nonumber
\\
	&-
	\frac{a_s^{\beta\gamma}(\omega,\omega')}{{4\pi(\hbar v_F)^2 }-{g}\,a_s^{\beta\gamma}(\omega,\omega')}
\label{eqn:KS}
,
\\
K_g^{\beta\gamma}(\omega,\omega')
	&=
	\frac{a_g^{\beta\gamma}(\omega,\omega')}{{4\pi(\hbar v_F)^2 }-{g}\,b_g^{\beta\gamma}(\omega,\omega')}
\nonumber
\\
	&-
	\frac{b_g^{\beta\gamma}(\omega,\omega')}{{4\pi(\hbar v_F)^2 }-{g}\,a_g^{\beta\gamma}(\omega,\omega')}
.
\label{eqn:KG}
\end{align}
\end{subequations}
As the optical response of doped graphene is due to excitation of carriers close to the Fermi
level, the sum over Landau levels in \pr{eqns:ab_strain_loop} is dominated by only a few
terms around the filling factor $n_0=\lfloor E_F^2/E_1^2\rfloor$.
For our parameters, the main contributions are found to come from the level $n=n_0$ and $n=n_0+1$, and we only retain these two terms
in the numerical evaluation of \pr{eqn:sxy_resum_scalar_3}.

Within this approximation, the Faraday angle calculated from $\sigma_{xy}^s$ and $\sigma_{xy}^g$ is shown in \pr{fig:disorderPlot}, where the blue curve corresponds to
scalar disorder and the red one to gauge disorder.
Assuming a graphene sample with $\mu$=$10\,000$\,cm$^2$V$^{-1}$s$^{-1}$, we set $\tau_q$=$\tau_{\rm tr}/2$=7.5$\times$10$^{-14}$.
Comparing  with the
classical expression \pr{eqn:sxy_Drude_int} shown in \pr{fig:classicPlot}, 
we note that neither type of disorder changes the functional form of $\Theta_F(\Omega)$: 
The  frequency of maximal rotation remains slightly below $\omega_c$ of \pr{eqn:wc_def}, and hence is determined by the external magnetic field.
This is in good accordance with experimental observations, where the data can be captured well by \pr{eqn:sxy_Drude_int},
if $\tau$ is treated as a fitting parameter. 
In contrast, a naive model of gauge disorder introducing a constant effective
field $B_{\rm eff}=B\pm\Delta B$ for carriers in the $\V K$ and
$\V{K'}$ valley, respectively, would shift the effective resonance  frequencies
in \pr{eqn:sxy_Drude_int} to 
$\omega_c^{\pm}=v_F^2|e(B\pm\Delta B)| /|E_F|$.

Concerning the magnitude of the achievable rotation angle, \pr{fig:disorderPlot} shows a larger value
for scalar disorder  compared to gauge disorder,
which stems from the different contribution of the vertex corrections to
$\sigma_{xy}^s$ and $\sigma_{xy}^g$.
In particular, upon expanding \pr{eqns:Kres} as
\begin{subequations}
\label{eqns:Kexp}
\begin{align}
K_s^{\beta\gamma}(\omega,\omega')
	&=
	\frac{1}{4\pi (\hbar v_F)^2} (b^{\beta \gamma}_s - a^{\beta \gamma}_s)
\label{eqn:KSexp}
\\[.5ex]
	&+
	\frac{g}{8\pi^2 (\hbar v_F)^4} [(b^{\beta \gamma}_s)^2 - (a^{\beta \gamma}_s)^2]
	+
	\mathcal{O}(g^2)
\nonumber
,
\\
K_g^{\beta\gamma}(\omega,\omega')
	&=
	\frac{1}{4\pi (\hbar v_F)^2} (b^{\beta \gamma}_s - a^{\beta \gamma}_s)
	+
	\mathcal{O}(g^2)
,
\label{eqn:KGexp}
\end{align}
\end{subequations}
we note that in the case of gauge disorder, vertex corrections to the simple bubble-diagrams are absent at 
linear order in $g$.
The simple bubble-diagrams given by the $\mathcal{O}(g^0)$ terms of  \pr{eqns:Kexp} 
are known to reproduce the classical Drude formula of \pr{eqn:sxy_Drude_int}, with $\tau=\tau_q$.
(See \pr{app:bubble} for details.)
Consequently, \pr{fig:disorderPlot} shows the maximal rotation angle in a sample with gauge disorder 
to be well aproximated by $\Theta_F^{\rm max}(\tau_q)$,
whereas the classical $\Theta_F^{\rm max}(\tau_{\rm tr})$ is restored by the contribution of vertex corrections
for scalar disorder\footnote{%
Ref.~\protect\onlinecite{Briskot_2013} shows for the case of $\sigma_{xx}$ in the presence of scalar disorder, 
how a resummation  of self-energy insertions and vertex corrections approximately restores the appearance of the transport scattering time
$\tau_{\rm tr}$ in \pr{eqn:sxy_Drude_int}}.

\begin{figure}
\centering
\includegraphics[width=0.85\columnwidth]{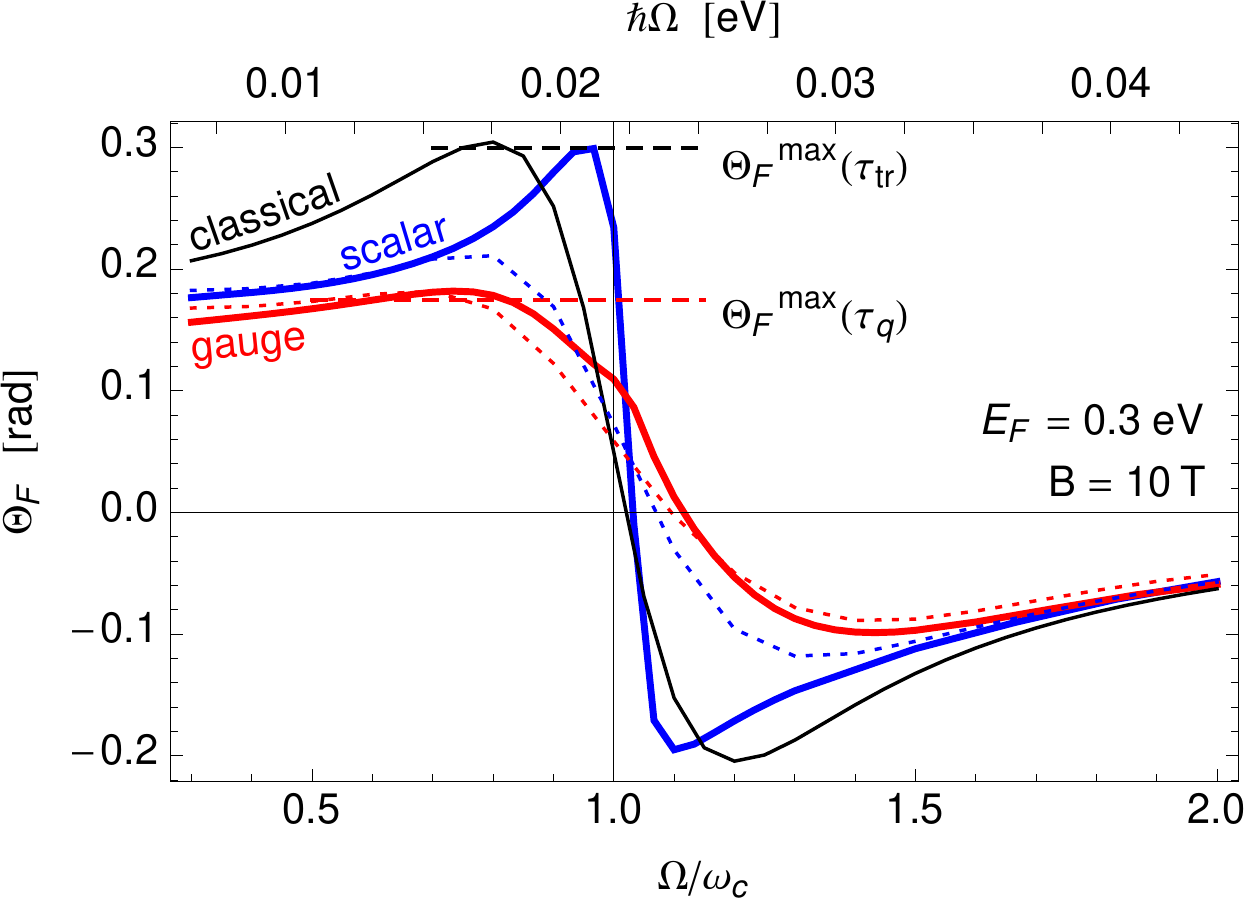}
\caption{%
Faraday rotation angle $\Theta_F$ 
calculated from the Kubo formula Eqn.~(\ref{eqn:sxy_resum_scalar_3}),
for a graphene sample with $\mu=10\,000$\,cm$^2$V$^{-1}$s$^{-1}$ due to scalar- (blue curve) and gauge disorder (red curve), respectively.
For comparison, the black curve re-displays the classical model shown in \pr{fig:classicPlot}.
The scattering time $\tau_q$ in \pr{eqn:Pidef} has been set to 7.5$\times$10$^{-14}$\,s, resulting in a disorder strength $g=0.058\,\hbar v_F$.
Horizontal dashed lines indicate the maximum angle given by \pr{eqn:max}.
The dotted curves show the results obtained by using only terms up to linear order in $g$ in the integral kernel  \pr{eqns:Kexp} 
of the Kubo formula.
\label{fig:disorderPlot}
}
\end{figure}

\section{Conclusions}
\label{sec:conclusions}
The dynamics of charge carriers which simultaneously experience the presence of a pseudomagnetic 
field $\Delta B$ as well as an external bias field $B$ is governed by an overall effective
field $B\pm\Delta B$, depending on whether the carriers reside near the $\V K$ or $\V{K}'$ point 
in graphene's Brillouin zone. 
This results in a different spectrum of Landau levels for $\V K$
and $\V{K}'$ carriers\cite{Roy_2013b,Low_2010},  and, in terms of the classical Hall conductivity 
$\sigma_{xy}$ of \pr{eqn:sxy_Drude_int}, one might therefore expect carriers at $\V K$ and $\V{K}'$ to 
contribute to the Faraday effect with a an effective cyclotron frequency of $\omega_c^{\pm}$,
respectively.
However, instead of resulting in an approximately constant pseudomagnetic field over an extended area,
the effect of sample-inherent ripples rather corresponds to  the exposure of graphene to a random
magnetic field, varying in direction and magnitude on a length scale of $\simeq$100\,\AA\cite{Meyer_2007}.

The above calculation of $\sigma_{xy}$, which regards the random pseudomagnetic field as a sublattice-mixing disorder potential, predicts that the frequency dependence of $\Theta_F$ is similar for both scalar- and gauge-
disorder mechanisms. Consequently, experimental data on Faraday rotation can be fitted well by assuming either scalar or gauge disorder,  if an appropriate $\tau$ is used as a phenomenological fitting parameter.
However, we suggest that in case the quantum scattering time $\tau_q$ of the specific sample is known,
the magnitude of the Faraday rotation might allow to differentiate between
samples with predominantly gauge disorder
and those showing predominantly scalar disorder.
An experimental possibility would be to determine $\tau_q$ by spectroscopic 
observation of the width of the relevant Landau levels near the Fermi energy\cite{Neugebauer_2009}, and compare it with the scattering time extracted form the maximum Faraday rotation observed in the same sample.

While  our calculations do not predict a shift of the observable effective  $\omega_c$ due to random strain fields,
we wish to emphasize that this does not exclude the possibility to influence the magneto-optical properties of graphene with  appropriately engineered macroscopic strain configurations.
\begin{acknowledgments}
The research leading to these results has received funding from the European
Union Seventh Framework Programme under grant agreement 604391 Graphene
Flagship, from the European Research Council through grant 290846, and from the
Spanish Ministry of Economy and Competitiveness under projects {MAT2014-53432-C5-1-R} and {FIS2014-207432}.
The authors are grateful for the hospitality of the Donostia International Physics Center (DIPC, San Sebastian),
where part of this work was concluded.
\end{acknowledgments}
\appendix
\section{Electron propagator and wavefunctions}
\label{sec:app_propagator}
The retarded (advanced) propagator for two-dimensional charge carriers in graphene reads\cite{Gusynin_2006,Miransky_2015}
\begin{align}
G^{\beta}(\omega,\V k,\tau)
	&=
	e^{-\V{k}^2 l_B^2}
	\sum_{n=0}^\infty
	\frac{(-1)^n G_n^{\beta}(\omega,\V k, \tau)}{[\hbar\omega + i \beta\hbar/(2\tau) ]^2-E_n^2}
,
\label{eqn:Gdef}
\end{align}
where $\beta=\pm1$ and $E_n$ denotes the energy of the $n$th Landau level, see \pr{eqn:LL_def}.
The numerator of \pr{eqn:Gdef} is given by
\begin{align}
G_n^{\beta}&(\omega,\V k, \tau)
	=
	2[\hbar\omega + i \beta\hbar/(2\tau )]\gamma^0
\nonumber
\\
	&\times
	\biggl\{
	L^{(0)}_n\bigl( 2 \V{k}^2 l_B^2  \bigr)
	\bigl[1 
	- i \gamma^1\gamma^2 \operatorname{sign}(e B) \bigr]/2
\nonumber
\\
	&-
	(1-\delta_{n,0})L^{(0)}_{n-1}\bigl( 2 \V{k}^2 l_B^2  \bigr)
	\bigl[1 
	+ i \gamma^1\gamma^2 \operatorname{sign}(e B) \bigr]/2
	\biggr\}
\nonumber
\\
	&
	- (1-\delta_{n,0})4 \hbar v_F \V{k}\cdot\vec{\gamma}
	L^{(1)}_{n-1}\bigl( 2 \V{k}^2 l_B^2  \bigr)
.
\label{eqn:GNdef}
\end{align}

The associated Laguerre polynomials $L_n^{(\alpha)}(x)$ appearing in \pr{eqn:GNdef} fulfill the orthogonality relation\cite{Gradshteyn}
\begin{align}
\int_0^\infty dx \,
x^\alpha e^{-x} L_n^{(\alpha)}(x) L_m^{(\alpha)}(x) 
	&=
	\frac{\Gamma(n+\alpha+1)}{n!}
	\delta_{n,m}
.
\label{eqn:otho_L}
\end{align}
Together with the identity
\begin{align}
L^{(m-n)}_n (x)
	&=
	\frac{m!}{n!} (-x)^{n-m} L^{(n-m)}_m (x)
,
\label{eqn:L_shift}
\end{align}
see Ref.~\cite{Gradshteyn},
\pr{eqn:otho_L} allows to compute 
the momentum integrals appearing in \pr{eqn:Kubo_Kernel_clean} and \pr{eqn:vertex_correction_general} of the main text.

The presence of Laguerre polynomials in \pr{eqn:GNdef} results from overlap integrals of Hermite polynomials $H_n(x)$,
\begin{align}
\int dx e^{-x^2} & H_m (x+y) H_n (x+z)
	=
\label{eqn:Hermite_to_L}
\\
	&
	\sqrt{\pi} 2^n m! z^{n-m} L^{(n-m)}_m (-2 y z)
	\qquad {\rm for} \quad m\leq n
\nonumber
.
\end{align}
The Hermite polynomials, in turn, appear in the electron wavefunctions in the presence of a magnetic field, see Ref.~\cite{Zheng_2002}.
\section{Evaluating bubble diagrams}
\label{app:bubble}
The simple bubble-diagrams treated in \pr{subsec:bubble} allow, apart from numerical evaluation of 
\pr{eqn:sxy_bare}
with the kernel $K_n^{(0)}$ of \pr{eqn:Kubo_Kernel_clean2}, a further analytical treatment:
One of the frequency integrations in \pr{eqn:sxy_bare} can be performed noting that,
by construction, $K_n^{(0)}$ is a purely imaginary function of $\omega$ and $\omega'$,
and fulfills $K_n^{(0)}(\omega,\omega')=-K_n^{(0)}(\omega',\omega)$,
such that we can rewrite \pr{eqn:sxy_bare} as
\begin{align}
\sigma_{xy}(\Omega)
	&=
	\frac{e^2}{h} \operatorname{Im}	\frac{4 \pi v_F^2}{\Omega}
	\sum_{n=0}^\infty
	\int d\omega n_F(\omega)
	\mathcal{K}_n^{(0)}(\Omega,\omega)
,
\label{eqn:sxy_bare2}
\end{align}
where
\begin{align}
\mathcal{K}_n^{(0)}(\Omega,\omega)
	&=
	\int \frac{d\omega' \, K_n^{(0)}(\omega,\omega')} {\omega-\omega'+\Omega+i 0}
	-
	\int \frac{d\omega' \, K_n^{(0)}(\omega,\omega')} {\omega-\omega'-\Omega-i 0}
.
\label{eqn:sxy_bare2a}
\end{align}
The $d\omega'$ integration in \pr{eqn:sxy_bare2a}
can be solved by closing
the contour in the complex $\omega'$-plane, 
the integrals that appear are of the form
\begin{align}
\int \frac{d\omega'}{2\pi i}&\frac{\omega' \pm i/(2\tau)}{[\omega-\omega' \pm (\Omega+i 0)] [(\omega'\pm i/(2\tau))^2-E_n^2]}
\nonumber
\\
	&=
	\frac{\Omega\pm\omega+i/(2\tau)}{E_n^2-[\Omega\pm\omega+i/(2\tau)]^2}
.
\nonumber
\end{align}
With $\mathcal{K}_n^{(0)}$ given by an  analytical expression, $\sigma_{xy}$ of \pr{eqn:sxy_bare2} 
can then be evaluated by numerically performing the remaining frequency integration and
truncating the sum over Landau levels at a suitable cut-off value. 

In the classical regime,
the spacing between single Landau levels at the Fermi energy is much
smaller than the Fermi energy itself, and
$
( E_{n_0+1} - E_{n_0}  )^2
	\approx
	(\hbar \omega_c)^2
$.
In this limit, 
one can proceed
to approximate the integrand in \pr{eqn:sxy_bare2}
and perform the remaining frequency integration. 
This yields 
the classical Drude formula of \pr{eqn:sxy_Drude_int}, with $\tau=\tau_q$, plus an
additional interband term which is non-resonant for $\Omega\ll2 E_F$.
For a detailed presentation of the calculational steps,  
we refer the reader to Ref.~\onlinecite{Gusynin_2007c}.
\end{document}